\documentclass[5p]{elsarticle}

\usepackage{natbib}
\usepackage[colorlinks]{hyperref}
\usepackage{amsmath}
\usepackage{xfrac}
\usepackage{enumerate}
\usepackage[nameinlink]{cleveref}

\crefname{figure}{Fig.}{Figs.}
\Crefname{figure}{Fig.}{Figs.}

\begin{document}

\begin{frontmatter}

\title{The Lukashian Calendar: A universal calendar mechanism using variable-duration years and days for any human colony or settlement}

\author{Gert-Jan Schouten}
\author{Michael Beye}

\begin{abstract}
Existing calendars, in particular the Gregorian calendar, have a complex structure which includes many adjustments in the
form of leap days and leap seconds in order to remain synchronized to the position of a planet in the Solar System. In addition, mean solar days can deviate up to 16 minutes
from true solar days on Earth. On other planets, this discrepancy can be far greater: on Mars, it can be up to 51 minutes. Numerous Martian proposals ranging from Aitken (1936)
to Gangale (1986) to Silin (2025) mimic the Gregorian calendar and do not address these issues. Moreover, many do not include a system for telling the time of day, or stretch the second beyond the standard
SI second. This paper proposes a calendar mechanism that uses years and days with a variable duration, in order to accurately represent true solar years and days. In this
mechanism, time of day is expressed as a proportion, rather than a number of fixed units, such as hours and minutes. This results in a simple and consistent structure that
can universally be applied to any planet, moon or space station with a human presence. Furthermore, it allows analog missions on Earth to use the same calendar mechanism as
future missions in space. This paper will outline the principles and details of this mechanism, using Earth and Mars as illustrative examples. Also discussed is the implementation
of the mechanism on actual computers, which have a Unix time system clock based on discrete milliseconds.
\end{abstract}


\end{frontmatter}

\section{Introduction}

\subsection{The Gregorian calendar}

The mechanism of the Gregorian calendar and the associated UTC time standard is the result of many centuries of development
and adjustment~\cite{BARNES1974}. It contains elements that are of astronomical origin, such as years and days, as well as elements with a cultural
or historical origin, such as hours, minutes or the seven-day week~\cite{MICHAELIS1982}. A summary of the main elements~\cite{HIGGINS2004,LOMBARDI2007,RICHARDS2013},
along with their numerical ranges, is as follows:

\begin{itemize}
\item Years, with a duration that approximately matches that of a solar year. The first year is 1582, the year in which the Gregorian calendar succeeded the Julian calendar.
	  Years before 1582 are proleptic, i.e.\ calculated backwards.
\item Months, with a duration that approximately matches that of a synodic lunar cycle and a range of 1--12 within a year.
\item Weeks, with a range of 1--52 or 1--53 within a year, depending on which day of the week a given year happens to start on.
\item Days, with a duration that approximately matches that of a mean solar day, a range of 1--7 within a week and a range of 1--28, 1--29, 1--30 or 1--31 within a month.
	  A leap day is added to the second month when the year is divisible by 4, unless it is also divisible by 100, but again if it is also divisible by 400.
\item Hours, with a range of 0--23 within a day, which becomes non-continuous in case of daylight saving time shifts.
\item Minutes, with a range of 0--59 within an hour.
\item Seconds, with a range of 0--59 or, in case of leap seconds, 0--60 within a minute. Seconds are also the official SI unit of time~\cite{IBWM9}.
\end{itemize}

Daylight saving time further complicates matters, with shifts toward the past being especially problematic. When the clock is turned back one hour, every timestamp within
that hour will occur twice: once before the clock is turned back, and then again after the clock is turned back. Correlating a timestamp to a unique point on the timeline
will then require extra information in addition to the timestamp itself. Turning the clock forward causes a non-continuous gap on the timeline~\cite{NISTDST2026}.

This already convoluted mechanism is further compounded by the use of time zones, along with their historical mergers, which caused the affected
regions to experience a one-off, non-continuous jump in the timeline (either a gap or a span of time that occurred twice). These must forever be taken into account when
performing calendar calculations involving timestamps that occurred before such mergers took place~\cite{OBS1909}.

\subsection{Mean solar days}

The use of days with a duration of approximately that of a mean solar day inevitably implies a difference between apparent
or true solar time and mean solar time, or calendar time. On Earth, this difference varies throughout the year from approximately --14 minutes to +16 minutes, meaning that the time of
day that one can deduce by observing the Sun, can be desynchronized anywhere from --14 minutes to +16 minutes with respect to the time of day on a Gregorian/UTC
clock~\cite[Chapter 28]{MEEUS1998}. This discrepancy is not ideal, but it is small enough to not significantly influence the daily routine of the average person.

On Mars, due to the much longer duration of a solar year, the fluctuations are considerably larger, ranging from --51 minutes to +40 minutes throughout a Martian
year~\cite{ALLISON1997}. This might be large enough for a person to experience in their daily routine, or to affect a person's circadian rhythm. This substantially
larger discrepancy between apparent solar time and calendar time implies that a calendar mechanism that uses mean solar days is much less suitable for Mars than it is for Earth.
On other planets, the discrepancy may be larger still.

Due to the slowing rotation of Earth, the duration of a mean solar day itself drifts over time. To address this, the Gregorian/UTC mechanism used to add leap seconds to
certain minutes of certain days~\cite{NELSON2001}. This has the unintended side effect that, whenever a second is added to a certain day, the duration of the year that
contains that day will also increase by one second. This is because, in the Gregorian/UTC mechanism, a year is constrained to span exactly a whole number of days.
From 2027, additional leap seconds will no longer be added~\cite{GCWM28}, causing calendar days to potentially drift from the solar cycle. Existing leap seconds
however, will forever have to be taken into account, as they will remain on the timeline.

\subsection{A new proposal}

In summary, the Gregorian/UTC calendar mechanism has a complex structure, with many elements that do not have an origin in nature and a timeline that is not continuous. The numerical
ranges of the various elements are inconsistent and affected by several corrective measures, such as daylight saving time, time zone mergers, leap days and leap seconds. Furthermore,
the discrepancy between apparent solar time and calendar time might not be too significant on Earth, but can be significantly larger on other planets, making
a calendar based on mean solar days intractable.

Current calendar proposals for Mars have mechanisms similar to the Gregorian/UTC mechanism. Examples include proposals by Aitken~\cite{AITKEN1936}, Moore~\cite{MOORE1977},
Gangale~\cite{GANGALE1986,GANGALE2006}, Šuráň~\cite{SURAN1997} and Silin~\cite{SILIN2025}. These proposals all involve mean solar days, years with a whole number of days, months, weeks and corrective measures,
such as leap days and leap seconds. Many do not include a system for keeping the time of day, or use a system that scales seconds, minutes and hours to fit a Martian
mean solar day, causing these seconds to no longer match the standard SI second~\cite{GANGALE2005}.

This paper proposes a mechanism called the Lukashian calendar. Employing elements with an origin in nature, it achieves astronomical accuracy by giving days and years variable durations
that match true, or apparent, solar years and days. This not only eliminates the discrepancy with mean solar days, but also avoids the need for leap seconds and leap days. Moreover, it
leads to a mechanism that can be universally employed to implement a calendar for any planet, moon or space station with a human presence.

\section{The Mechanism}

\subsection{Instances with Years and Days}\label{INSTANCES}

Each location, such as a planet, moon or space station that uses the Lukashian calendar has its own separate instance of it. An instance of the Lukashian calendar defines the years,
days and epoch\footnote{In the context of calendars, the term “epoch” refers to a point in time at which a calendar starts.} for that particular location and has the following properties:

\begin{itemize}
	\item The durations of years and days (expressed in milliseconds) are variable: one year or day can have a different duration than the next.
	\item A year does not necessarily have a whole number of days. A day belongs to the year in which it starts, even if it ends in the next year.
	\item The first year is year 1 and years before year 1 are not defined.
	\item The first day of each year is day 1 and each following day of the year is sequentially numbered.
	\item The first year and the first day of the instance start at the epoch of the instance.
	\item Years and days have an inclusive start and an exclusive end. This effectively implies that the instance has a start (the epoch), but not an end.
	\item The Lukashian calendar does not have time zones, the turn of day is the same across the entire location where the instance is in use.
\end{itemize}

%These images should be placed in the latex source such that they appear at the top of page 3; the exact location depends on the chosen font settings
\begin{figure*}
    \centering
    \includegraphics{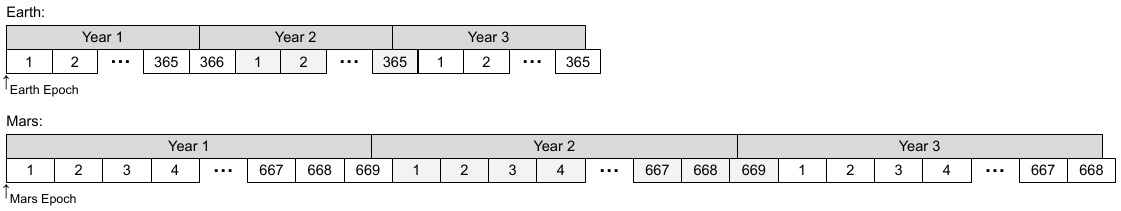}
    \caption{
		Schematic of the first three years of the Earth and Mars instances, including their days.
		Note: the instances are visually left-aligned in this figure, but have distinct epochs, which do not coincide.
	}\label{FIRST3YEARS}
\end{figure*}
\begin{figure*}
    \centering
    \includegraphics{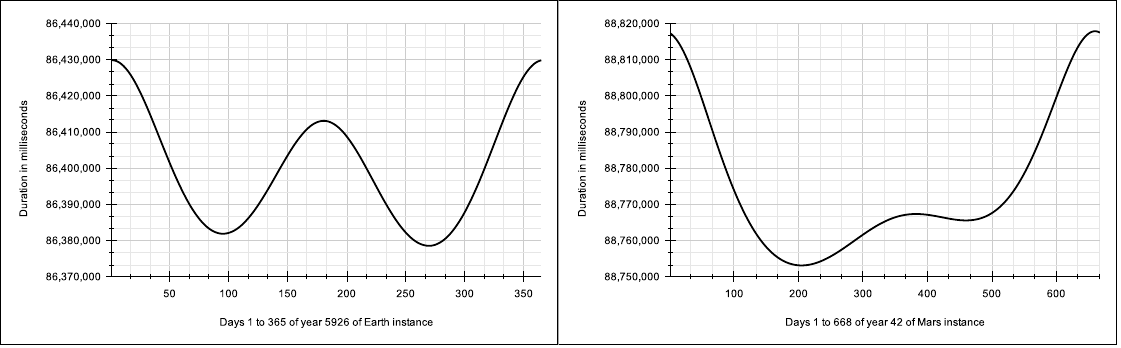}
    \caption{Charts demonstrating the variable duration of Lukashian days.}\label{DAY_DURATIONS}
\end{figure*}

A simple case would be an instance of the Lukashian calendar for an intergenerational starship to another solar system. Such a location has no constraints following from a planetary orbit
or rotation. Years and days can therefore be freely defined as one deems convenient. One could define all days to have a duration of exactly 86,400 seconds, to match the human circadian
rhythm, and all years to have a duration of exactly 400 days, in order to have a round number that is easy to work with. Since the option of having years and days with a variable duration
is not used, each year will conveniently have a whole number of days.

An instance for a planet such as Earth or Mars, however, is constrained by the orbit and rotation of the planet. Years should match the seasonal cycle following from the planet's orbit and
days should match the day-night cycle following from its rotation. Variable duration years and days are used to satisfy these constraints.

On Earth and Mars, a calendar year is defined as a solar year that runs from one southern solstice (inclusive) to the next southern solstice (exclusive), exactly. The duration of a solar
year is not constant. On Earth, the duration of one particular solar year can be different from the duration of another by hundreds of seconds~\cite{RICHARDS2013}.

On Earth and Mars, a calendar day is defined as a true, or apparent, solar day, from its start (inclusive), to its end (exclusive). Due to the eccentricity of the orbits of the
planets, as well as their obliquity, the duration of a true solar day is not constant. On Earth, the difference between the shortest and longest true solar day of any particular
year is approximately 51 seconds~\cite[Chapter 28]{MEEUS1998}. On Mars, it is approximately 65 seconds~\cite{ALLISON2000}.

These definitions imply that there is no whole number of calendar days per calendar year, as the yearly cycle of the seasons is essentially independent of the daily cycle.

\Cref{FIRST3YEARS} shows a schematic of the first three years of the Lukashian calendar instances for planets Earth and Mars. A Martian true solar day, or sol, is
approximately 2,375 seconds longer than a true solar day on Earth~\cite[Chapter 28]{MEEUS1998}~\cite{ALLISON2000}. A Martian year, however, is much longer than a year on Earth,
with each Martian year containing 668 or 669 Martian days. The days are numbered 1 to 365 or 366 in the Earth instance of the Lukashian calendar and 1 to 668
or 669 in the Mars instance. The combination of a year number and a day number uniquely identifies a day of an instance. For example, the date “253 -- 5926” denotes day 253 of
year 5926 in the Earth instance.

The universal nature of the mechanism can be observed from \cref{FIRST3YEARS}. Indeed, the mechanism makes no assumptions about the duration of any particular
year or day, or the number of days in a year. On Earth and Mars, true solar years and days are used. For a moon, space station, or the aforementioned intergenerational space ship
to another solar system, entirely different definitions of a year and a day may be chosen. For habitable planets with very short,
or very long rotational periods, instances may opt for day durations that do not match those periods, since the day-night cycle of humans on those planets will,
for biological reasons, not match those periods anyway. The same principle applies to planets with very long or very short orbital periods. Year durations may be chosen differently
from those periods. Each instance must determine the durations of years and days according to the specific characteristics of the location and the needs of its residents.

The purpose of the Lukashian calendar is not to provide one date and one time across multiple locations in space, since such a mechanism would be unrelated to the seasonal and
day-night cycles of individual locations and thus inconvenient for their residents.

What can also be observed from \cref{FIRST3YEARS}, is that certain years will automatically gain an extra day. This no longer requires complicated, artificial rules for
when to add leap days.

For Earth, the reason for starting a year at southern solstice is twofold. First, southern solstice is not too far from New Year in the Gregorian/UTC calendar, which is
convenient for adoption and societal transition. Second, southern solstice is a natural point in time to start a year. Other natural points, such as northern solstice or one of
the equinoxes, are much further from Gregorian/UTC New Year. For Mars, southern solstice was chosen in order to be consistent with Earth. Note however, that the calendar
mechanism makes no assumption about when a year should start. Should consensus decide on a different starting point for a Martian year, then the instance for Mars is easily adjusted
(prior to adoption, of course).

\Cref{DAY_DURATIONS} shows the durations of the days of year 5926 of the Earth instance and year 42 of the Mars instance.\footnote{These particular years were chosen
because they were the current years at the time of writing of this paper, given the choice of epochs as explained in \cref{CHOOSING_THE_EPOCH}.} These accurately match the durations
of true solar days. For verification, Fig. 1 from Allison~\cite{ALLISON1997} could be observed. Please note that Allison starts the Martian year at vernal equinox, instead of southern
solstice, so the dashed line should be moved rightward by a quarter before comparing.

By eliminating the discrepancy between true and mean solar days, any possibility of this discrepancy affecting a person's daily routine or circadian rhythm is avoided.
Also, the addition of leap seconds to adjust for the drift of a solar day is no longer needed.

\subsection{Weeks}\label{WEEKS}

A Lukashian week is defined as a period of 10 days, with the first 7 being working days and the last 3 being weekend days. Three days out of ten results in a ratio of 0.3, whereas two days
out of seven equals 0.286, making a Lukashian weekend on Earth only slightly longer than a Gregorian/UTC one. A duration of 10 days makes calculations easy, since society at large
uses the decimal system.

\begin{figure}
\centering
\includegraphics{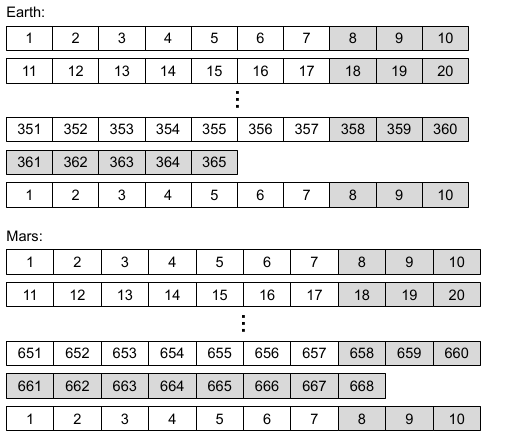}
\caption{Schematic of the weeks and day numbers of typical years in the Earth and Mars instances, with weekend days and the final week of the year in gray.}\label{WEEK_NUMBERS}
\end{figure}

\Cref{WEEK_NUMBERS} shows a schematic of the weeks and day numbers of typical years in the Earth and Mars instances. The first week of any particular year starts on day 1
of that year. On Earth and Mars, this requires the final week of a year to have less than 10 days. If a particular year in the Earth instance has 366 days instead of 365,
or 669 instead of 668 in the Mars instance, then the final week will contain the extra day.

Many cultures worldwide celebrate festivities toward the end of the year. The final, shortened week of the year could therefore be designated a holiday week of
sorts. In fact, New Year could be celebrated on the final day of the year, as the next year will, in actual fact, start during that day, as can be observed from
\cref{FIRST3YEARS}.

The Lukashian calendar does not have months. The synodic lunar cycle, with a duration of approximately 29.5 days~\cite{RICHARDS2013}, has little influence on the lives of the
average person and does not accurately match a Gregorian/UTC month in the first place. On Mars, with its two small moons, the synodic lunar cycles are even less relevant to the
inhabitants, so the concept of a month is not essential and would only add complexity to the calendar mechanism. Moreover, most, if not all, subdivisions of a year into months
(e.g.\ the Gregorian/UTC calendar or Gangale's Darian calendar proposal for Mars), do not manage to get a whole number of weeks, or a constant number of days, into a single month.
As the Lukashian calendar aims to provide a universal mechanism for every planet, moon, space station or otherwise, the concept of months simply does not fit.

%This image should be placed in the latex source such that it appears at the top of the right column of page 4; the exact location depends on the chosen font settings

%This image should be placed in the latex source such that it appears at the top of page 5; the exact location depends on the chosen font settings
\begin{figure*}
\centering
\includegraphics{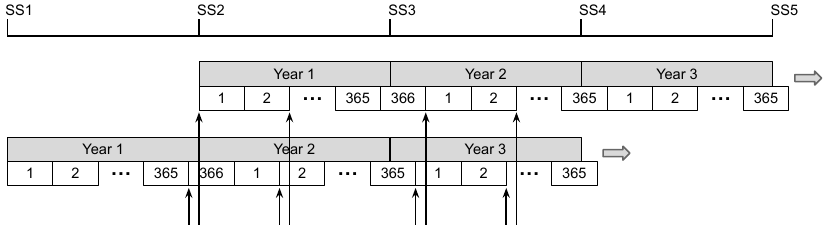}
\caption{The effect of the chosen epoch on the turn of day.}\label{EPOCH}
\end{figure*}

Additionally, weeks with a duration of 10 days uniquely provide more structure than weeks with a different number of days. This is because the decimal system will conveniently
indicate which week of the year corresponds to any given day number. For example, day 253 is part of the 26\textsuperscript{th} week of the year. One simply needs to look at
the leading digits of the day number (25) and add 1. A calendar mechanism with a 7-day week does not allow this, especially when the first week does not necessarily start on
the first day of the year, as is the case with the Gregorian/UTC calendar. This ability to easily deduce a week's position within the year, merely by observing the day number,
offsets the absence of months in the Lukashian calendar.

A 10-day week eliminates the need for naming weekdays. The Gregorian/UTC calendar uses weekday names such as “Monday” and “Tuesday”, whereas the Darian calendar proposal
for Mars uses names such as “Solis” and “Lunae”~\cite{GANGALE1986,GANGALE2006}. Using the Lukashian calendar, one could simply look at the last digit of the day number in order
to know which day of the week it is. For example, day 253 is the 3\textsuperscript{rd} day of its week. It is easy to imagine an alternative to the use of weekday names in everyday
language. In English, for example, one could use the term “Third-day” to denote the third day of a week, as in: “I'm having an important meeting on Third-day this week”.
This would work equally well for day 251 (“First-day”) up to day 260 (“Tenth-day”).

\subsection{Choosing the Epoch}\label{CHOOSING_THE_EPOCH}

As stated, each instance of the Lukashian calendar, for example the Earth instance, starts at a particular point in time: the epoch.
The first year, as well as the first day of that particular instance, will start at the epoch. This is illustrated in \cref{FIRST3YEARS}. For instances
of the Lukashian calendar that are constructed for planets, choosing the epoch is the only thing that can be decided upon. After choosing the epoch, the laws of astronomy will
determine the remainder of the course of the calendar: years will be true solar years and days will be true solar days.

If southern solstice is chosen as the starting point of a year, then a particular southern solstice needs to be chosen as epoch.
After all, the first year will start at the epoch. In order to determine which southern solstices are most suitable as epochs of the Earth and Mars instances,
one needs to consider the effect that the epoch has on the turn of day (the point at which one day ends and the next day starts).

Because the Lukashian calendar has no time zones, note that the turn of day will happen simultaneously across the planet.

\Cref{EPOCH} presents five hypothetical southern solstices, labelled \textit{SS1} to \textit{SS5}. It also presents two hypothetical Lukashian calendar instances that are constructed
by choosing \textit{SS2} respectively \textit{SS1} as epoch. It can be observed that the years of both instances run from one southern solstice to the next.
The days however, show a significant difference between the two instances when it comes to their starting points, i.e., the turn of day.

This difference is caused by the fact that the epoch defines when every year starts, as well as when every day starts. The epoch is likely the only point in time when a
year starts at the same point as a day. It is also the only thing that can be chosen, after which the laws of astronomy will take over. This means that, if there
is a preferred point in time to start a year (e.g.\ southern solstice), and a preferred point in time to start a day (which will be the same across the planet),
then the epoch must coincide with both.

The preferred point in time to start a day is when it is nighttime for the majority of the population on the planet in question. This means that the region that holds
the majority of the population must face away from the Sun at the epoch.

For Earth, the southern solstice that was chosen to be the epoch is the one that occurred in the Gregorian year 3900 BC\@. The current Gregorian year of 2026 (at the time of writing of
this paper), would then largely overlap with Lukashian year 5926. This particular southern solstice was chosen because it satisfies all three of the following conditions:

\begin{enumerate}
\item The last two digits of the year match the last two digits of the Gregorian year, which is convenient for adoption and transition. The first 10 days or so of a new Lukashian Earth
	  year are an exception to this rule, as southern solstice takes place approximately 10 days before the end of a Gregorian year.
\item Human written history started approximately 5000 years ago~\cite{WALKER1987}. Choosing an epoch at least 5000 years ago therefore ensures that all events with any sort
	  of date written in human history “fit” in the calendar, since, in the Lukashian calendar mechanism, years before year 1 are not defined.\\
	  Events that happened before this, generally do not have a known date and are referred to by number of years ago, for example: “The end of the last glacial period
	  was 12,000 years ago”. Such prehistoric events are not referred to by calendar date, so a calendar need not accommodate those.
\item The region from Western Europe / West Africa to East Asia / Australia experiences the turn of day around nighttime. Approximately 87\% of Earth's population lives in
	  this region~\cite{WORLDOMETERS}. The turn of day will happen during the early morning in the East of this region and during the late evening in the West of it.
\end{enumerate}

For Mars, the southern solstice that was provisionally chosen to be the epoch is the one that occurred on Gregorian date April 27, 1947. This means that the current Martian year is 42
at the time of this writing, since Mars has completed 42 orbits around the Sun since the mentioned date~\cite{ALLISON2000}. This particular southern solstice was chosen
by evaluating the same three conditions for planet Mars:

\begin{enumerate}
\item The convenience of the last two digits of the year matching an existing calendar does not apply, since there is no existing calendar on Mars.
\item Human history on Mars started on Gregorian date July 15, 1965 with the Mariner 4 flyby~\cite{MARINERMEMORANDUM}, so accommodating all of human history on Mars requires
	  the epoch to be at a point before that particular date.
\item Since the location of a future settlement is not yet known, the region which must face away from the Sun at the epoch cannot yet be established. In order to construct
	  a provisional Lukashian calendar instance for planet Mars, the site of the Pathfinder lander was chosen to act as a provisional settlement location. The choice for
	  Pathfinder was inspired by the fact that Pathfinder deployed the first ever successful rover~\cite{BAJRACHARYA2008}. At the chosen southern solstice, the Pathfinder
	  site was reasonably close to the longitudinal center of the region that faced away from the Sun.
\end{enumerate}

A consequence of choosing this particular epoch for the Mars instance is that all timestamps in the Earth instance that occur before the Mars epoch, do not have a corresponding
timestamp in the Mars instance. Should this be considered undesirable, an earlier epoch for Mars can be chosen. For example, the Martian southern solstice closest to the Earth epoch
can be chosen, subject to the availability of an algorithm to calculate this with sufficient accuracy.

\subsection{Time of Day}

Because days in the Lukashian calendar have a variable duration, the concept of “time of day” cannot work in the same fashion as a Gregorian/UTC clock does. Hours and minutes
have a fixed duration, so it is unlikely that any Lukashian day will have a whole number of hours or minutes. The same holds for any other unit of time.

For this reason, the Lukashian calendar does not express time of day using units, but rather as a proportion. The time of day is simply the proportion of the day that has
passed. To represent this proportion on a standard Lukashian clock, basis points, or permyriad, are used.  A basis point (“bp” or “beep”), is one hundredth of one percentage point,
or one ten-thousandth. For example, day 80 of year 5926 of the Earth instance of the Lukashian calendar has a duration of 86,384,256 milliseconds. When 30,000,000 of those milliseconds
have passed, the proportion of that day that has passed is 0.34728550535875424. This proportion would then be represented as “3472”. Note that the digits
after the first 4 decimal digits are truncated, not rounded up. This ensures a number between 0000 and 9999, making it represent the number of beeps that have fully passed.
Of course, if a higher precision than beeps is required, clocks with additional significant digits could be used.

In the Earth instance of the Lukashian calendar, one beep has a variable duration of approximately 8.6 seconds. This implies that, for intuitive use, the following durations are
roughly equivalent:
\begin{align*}
\text{10~beeps~}&\approx\text{~1\sfrac{1}{2}~minutes}\\
\text{100~beeps~}&\approx\text{~15~minutes}\\
\text{1,000~beeps~}&\approx\text{~2\sfrac{1}{2}~hours}\\
\text{10,000~beeps~}&=\text{~1~true~solar~day}
\end{align*}
In the Mars instance, a beep lasts approximately 8.9 seconds, because of longer solar days.

An example of a timestamp in the Gregorian/UTC calendar, using format “{\small [day], DD-MM-YYYY HH:mm:ss ZZZ}”, is:
\begin{center}
\text{Wednesday, 11--03--2026 06:24:59 CET}
\end{center}
The same point in time expressed in the Darian calendar proposal for Mars, is:
\begin{center}
\text{Sol Jovis, 26--17--221 09:47:44 Airy}
\end{center}
The same point in time expressed in the Earth instance of the Lukashian calendar, using format “{\small DD-YYYY BBBB}”, is:
\begin{center}
\text{80--5926 3472}
\end{center}
The same point in time expressed in the Mars instance of the Lukashian calendar is:
\begin{center}
\text{625--42 2486}
\end{center}

%This image should be placed in the latex source such that it appears at the top of page 7; the exact location depends on the chosen font settings
\begin{figure*}
\centering
\includegraphics{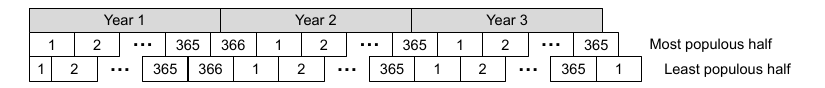}
\caption{Shifting the days for the least populous half of the planet.}\label{TIMEZONES}
\end{figure*}

Since days have a variable duration, beeps of different days have a different duration. In the Earth instance, for example, the shortest beep of a year is approximately 5 milliseconds
shorter than the longest. Because of this, beeps should mainly be used to express the time of day. Using beeps to express durations is possible, but only if low precision is sufficient
and small deviations are acceptable. For example, the phrase: “Travel time to work is 200 beeps” would be appropriate, but the phrase: “The runner finished the race in 153.442 beeps”
would not, because this amount of beeps would correspond to different actual durations on different days. In cases such as this, which are about duration, not time of day, durations
should be expressed in seconds, as the second is the official SI unit of time.

In the Gregorian/UTC calendar, as well as various calendar proposals for Mars, the second is also a calendar unit, since minutes, hours and days are all defined in terms of seconds.
This duality can lead to ambiguous situations, for example when leap seconds are present on the timeline, or when Martian calendar proposals stretch the second to make 24 hours match
a Martian mean solar day.

Compared to calendars that use hours, minutes and seconds to represent the time of day, the Lukashian calendar is not only simpler, but also provides more insight
into the proportion of one's time spent. For example, if someone spends 1500 beeps doing research, this directly translates to 15\% of that person's day.
Hours and minutes do not provide this immediate insight.

From a numerical point of view, the numbers 24 (hours) and 60 (minutes and seconds) are convenient to work with, due to their many divisors. The number 10,000 however,
has a convenient property of its own. It is a power of 10, which, like the 10-day week, is useful, considering society's decimal system.

The Lukashian calendar does not incorporate the concept of daylight saving time. Existing research suggests that the cost/benefit ratio of employing daylight saving time is
dubious at best~\cite{COSTAFONT2024}.

\subsection{An argument against time zones}

As mentioned before, the Lukashian calendar does not have time zones; the time of day, and therefore the turn of day, will be the same across a planet. This means that daily
schedules will differ. For example, a person in a certain region might wake up at 3300 beeps, go to work at 3700 beeps and come home at 7500 beeps, but another person living in
a region with a different longitude will have their entire daily schedule offset by several hundreds or even thousands of beeps. If these two people were to share an activity, for
example an online meeting, they would have to take this into account. The time of day would indicate the same point in time for both, but for one it might be nighttime, whereas
for the other, it might be early morning. In the Gregorian/UTC calendar, a timestamp always corresponds to the same phase of the day: 07:00 always indicates early morning. However,
07:00 indicates a different point in time for each person, which they would have to take into account instead. In the end, coordination between people living in different longitudinal regions
will always be necessary.

On Earth, the turn of day will be around nighttime for 87\% of the population~\cite{WORLDOMETERS}, which means that it will be during daytime for the
remaining 13\%. This is a concession made for the benefit of having the same time of day across a planet. It is possible to imagine a mechanism where the least populous half of
the planet has its sequence of days shifted by half a day, with the first day of the calendar having a duration of only half a day, as illustrated in \cref{TIMEZONES}.
This construct, however, would have the following problems:

\begin{itemize}
\item The days of the least populous half of the planet will be shifted, but the years should remain the same, since years must run from southern solstice to southern solstice. This implies
	  that days of the least populous half may no longer belong to the year in which they start.
\item The days of the least populous half can be defined to start at exactly 5000 beeps of the days of the most populous half, but the reverse would then not be true. Due to the variable
	  durations of days, the days of the most populous half will only approximately start at 5000 beeps of the days of the least populous half.
\item This mechanism would not follow directly from nature.
\item This mechanism adds complexity and potentially opens the door to not just two, but more “time zones”.
\end{itemize}

Once the population of, say, Mars is accustomed to a calendar with a single time of day across the planet, its people may actually prefer uniformity over having
the turn of day during nighttime. After all, the 12-hour Gregorian/UTC clock restarts the counting of hours at 12 noon, but this is not considered an inconvenience, because
everyone is used to it. The use of the mechanism outlined above, or any other mechanism that resembles time zones, is therefore not recommended.

\section{Computational Implementation}

\subsection{Continuous to Discrete}\label{CONTINUOUS_TO_DISCRETE}

%This image should be placed in the latex source such that it appears at the top of page 8; the exact location depends on the chosen font settings
\begin{figure*}
\centering
\includegraphics{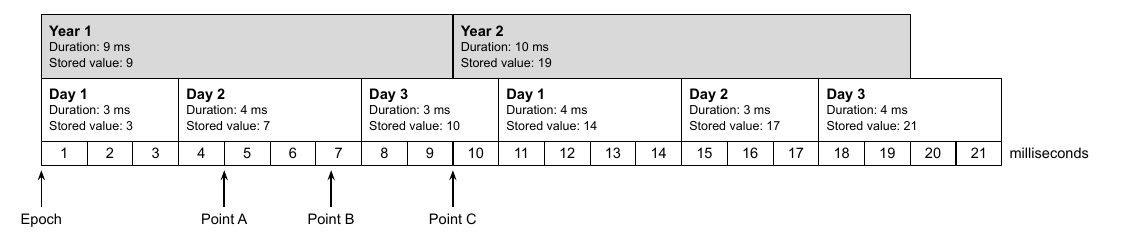}
\caption{Lukashian years and days projected onto a discrete millisecond timeline.}\label{TIMELINE}
\end{figure*}

The Lukashian calendar mechanism outlined so far is based on a timeline that is continuous. The reference implementation of the Lukashian calendar, written in the
Java programming language~\cite{REFIMPL}, projects this mechanism onto a computer's discrete timeline, usually consisting of whole milliseconds and using Unix time.

For each instance of the Lukashian calendar, the number of milliseconds between the Unix epoch and the Lukashian epoch is specified, enabling computers to efficiently translate a Unix
timestamp to the corresponding Lukashian timestamp and vice versa, as well as to convert a Lukashian timestamp in one instance to the corresponding timestamp in another instance
(provided of course, that the timestamp in question is defined in both instances). These translations compensate for any leap seconds present on the Unix timeline.

Then, starting at the Lukashian epoch, the durations, in milliseconds, of the years and days are calculated. For the Earth instance, this is done using Meeus' solstice
algorithm~\cite[Chapter 27]{MEEUS1998} for years and Meeus' equation of time~\cite[Chapters 22,25,28]{MEEUS1998} for days. These have an acceptable accuracy for timespans of several
thousands of years. Therefore, durations for 7000 years (and their days) after the Earth epoch are calculated. For the Mars instance, this is done using Allison \& McEwen's
paper for both years and days~\cite{ALLISON2000}. This paper specifies Martian solstices up to 96 Martian years after the Lukashian Mars epoch. Both Meeus and
Allison \& McEwen use Julian Days to calculate durations, implying that the durations of years and days of both calendar instances are calculated in terms of Terrestrial
Time~\cite[Resolution B1, Definition 2]{IAU1997} and with it, the current definition of the SI second~\cite[Resolution A4, Recommendation IV]{IAU1991}. In the future, additional
durations of years and days should be added when algorithms to calculate those with sufficient accuracy are available.

For each year and day, the implementation stores the number of milliseconds between the Lukashian epoch and the end of that particular year or day. This enables the mechanism
to determine the corresponding year and day, given a certain millisecond on the timeline, which is essentially an offset from the epoch. The duration of each year and day can
then be determined by subtracting its predecessor's value from its own. The very first year and day of the calendar have no predecessor.

\Cref{TIMELINE} shows an illustrative toy example: a fictional instance of the Lukashian calendar with years and days lasting only milliseconds. It is
projected on top of a discrete millisecond timeline, with the first millisecond having index 1. Points A, B and C are included for later reference.

\subsection{A Point in Time}

Most programming languages provide a data type that represents a unique point on the timeline, commonly with millisecond precision. In Java, this type is called an “Instant”. The terms
“DateTime” or “Timestamp” are also common.

A Lukashian Instant does not only represent a certain millisecond on the timeline. Since the Lukashian calendar expresses the time of day as the proportion of the day that has
elapsed, it also represents that very proportion. A programmer should be able to instantiate an Instant by specifying a certain millisecond ($m$), or by specifying a certain
proportion ($p$) of a certain day ($D$):

\begin{itemize}
\item Let $m$ be a certain millisecond on the timeline.
\item Let $D$ be the day that contains $m$. By \cref{CONTINUOUS_TO_DISCRETE}, if $m$ is known, $D$ is known.
\item Let $d$ be the total duration of $D$ in milliseconds. By \cref{CONTINUOUS_TO_DISCRETE}, if $D$ is known, $d$ is known.
\item Let $e$ be the number of milliseconds of $D$ that precede $m$.
\item Let $p$ be the proportion of $D$ that has elapsed.
\end{itemize}

If $m$ is specified, then the point in time that is considered to have been specified is the point at the start of $m$. In this case, $e$ and $d$ are known, thus $p$ can be computed:
$p = \sfrac{e}{d}$. For example, if $m$ is millisecond 5 in \cref{TIMELINE}, then point A on the timeline is considered to have been specified. $D$ is then day 2 of year 1
and $p = \sfrac{e}{d} = \sfrac{1}{4}$ (which would then be displayed as 2500 beeps on a standard precision Lukashian clock).

This approach implies that specifying the first millisecond of a
day effectively specifies the actual start of that day and will result in a proportion of 0. Specifying the first millisecond on the timeline effectively specifies the epoch.
It also ensures the correct implementation of the definition given in \cref{INSTANCES}, which states that days have an inclusive start and an exclusive end. Specifying
the last millisecond of a day does not effectively specify the end of that day, as the end of that day is actually the start of the next. Instead, it effectively specifies
the point in time one millisecond before the end of that day and will result in a proportion close to, but smaller than 1.

Since $e$ and $d$ are always integer numbers and $p = \sfrac{e}{d}$, $p$ is guaranteed to be a rational number. Many calculations involving Instants, such as adding a proportion,
or calculating the difference between two Instants, can be performed without any loss of precision. This is achieved by expressing the calculation in terms of $e$ and $d$, finding
a common denominator and then performing an integer addition or subtraction.

If $p$ and $D$ are specified, only rational numbers are accepted for $p$, in order to maintain precision. Then, $m$ is determined to be the millisecond that “contains” $p$.
For example, if $p = \sfrac{5}{6}$ and $D$ is day 2 of year 1 (point B on the timeline in \cref{TIMELINE}), then $m$ is millisecond 7. If $p$ happens to be exactly on the
boundary between milliseconds, then $m$ is the millisecond starting at the boundary. For example, if $p = \sfrac{1}{4}$ and $D$ is day 2 of year 1 (point A on the timeline),
then $m$ is millisecond 5. This is evident: if $m = 5$ leads to the conclusion that $D$ is day 2 of year 1 and $p = \sfrac{e}{d} = \sfrac{1}{4}$, then the reverse should also hold.
This ensures that days have an inclusive start and an exclusive end. In fact, the mechanism outlined above essentially implies that, in the computational implementation
of the Lukashian Calendar, individual milliseconds on the timeline also have an inclusive start and an exclusive end.

Although the concept of “time of day” does not apply to years, the same principle holds for the start and end of a year. For example, point C on the timeline in \cref{TIMELINE} is
part of year 2. Years, days and individual milliseconds are mutually consistent in that they start their count at 1, have an inclusive start and an exclusive end.

\subsection{Standardization and Maintenance}

Standardization of the durations of years and days is paramount for adoption and real world usage. The reference implementation provides a mechanism
for loading the durations from an external resource, for example a resource that is centrally maintained and governed by an official organization. This organization is then
responsible for choosing and improving the algorithms used for calculating the durations. Such an organization  may have reasons for changing the durations of years and/or days
for a certain instance. These changes can be made with different levels of invasiveness:

\begin{enumerate}
\item Replace the durations of all years and days. This can be done, for example, when consensus decides on a different epoch for a certain instance,
	  and should only be done prior to adoption.
\item Replace the durations of years and/or days that start after a certain offset from the present, for example 100 years. This can be done, for example, when a more accurate
	  algorithm for calculating the durations becomes available.
\item Add durations of years and days in the far future that aren't yet defined in the instance. This should be done whenever an algorithm for calculating those durations
	  with sufficient accuracy becomes available.
\end{enumerate}

It could happen that an algorithm which is believed to be accurate, turns out to be less so. For example, the length of a true solar day could increase faster than anticipated.
This situation can be addressed by adopting a new algorithm for calculating the durations of days, while leaving the algorithm for calculating
the durations of years unchanged. A crossover day $D_{co}$ in the future will have to be chosen. Every day after $D_{co}$ will be calculated using the improved
algorithm. The duration of $D_{co}$ itself can be chosen such that it will account for the one-off “correction” between the old and the new algorithm. The benefit of this approach is that
the calendar mechanism itself does not need to provide a dedicated construct for corrections and still simply requires a list of durations of years and days. The timeline
will therefore always remain continuous and unambiguous: it will never contain gaps or repeated timestamps.

\section{Analog Research}

The universal nature of the Lukashian calendar allows analog missions on Earth to use the same calendar mechanism as astronauts inhabiting a human settlement on, for example, Mars.
Analog astronauts on Earth can use the Earth instance, in order to remain synchronized with the day-night cycle on Earth, whereas astronauts on Mars use the Mars instance. The durations
of days and years will be different, but the numbers of the calendar will have the same meaning and therefore provide a comparable experience. This enables evaluation of the
Lukashian calendar by analog astronauts, as well as preparation and practice on Earth by astronauts traveling to Mars.

An analog mission could study how people experience a calendar containing only years, weeks, days and time of day, all expressed using only numbers, with the number of days
per week and the time of day optimized for the decimal system. This study could include an analysis of how long it takes for people to get accustomed to the Lukashian
calendar, given that the Gregorian/UTC calendar is the current convention. Of course, Mars settlers will inevitably have to get accustomed to some different sort of calendar in any case,
since the Gregorian/UTC calendar does not work on Mars.

To study how people experience a calendar with a single time of day across the planet, communication across analog stations covering different longitudinal regions could be
enabled. This permits analog astronauts to coordinate common schedules without having to take into account different time zones.

Furthermore, since the Lukashian calendar provides only numbers, it is worth analyzing how usage of the Lukashian calendar will develop from a linguistic point of view. Will
new language constructs or abbreviations be introduced? Will certain durations (e.g.\ 10, 100 or 1000 beeps) be designated with certain words? Will weekdays indeed be referred to
as suggested in \cref{WEEKS}? The outcome of such an analysis might be different across languages. Follow-up analysis could then compare.

Regarding the discrepancy between true and mean solar days, a comparative analysis could be performed at a time of year when this discrepancy is largest. Such an analysis
could compare analog astronauts' daily schedules created with the Gregorian/UTC calendar, against those created with the Lukashian calendar. On Earth, the maximum discrepancy
is 16 minutes, which is not as prominent as the maximum of 51 minutes on Mars. This should be taken into account when interpreting the results.

The North and South Pole of Earth experience only a single day and a single night per year, each lasting half of the year. Analog stations near the Poles therefore don't follow the
regular Earth day-night cycle. This gives such stations the opportunity to use the Mars instance of the Lukashian calendar for practice and training. Communication with analog
stations further from the poles (which would use the Earth instance) could be established in order to simulate real-world conversion between the Earth and Mars instances.

In addition to above examples, more psychological, technological, sociological and linguistic experiments could be performed. Further research is required to establish suitable
analog mission objectives.

\section{Conclusion}

Current calendar proposals for Mars are conceptually similar to the Gregorian/UTC calendar. They are affected by a discrepancy between true and mean solar days and have a complex
mechanism, requiring leap days and leap seconds in order to stay synchronized with planetary motions. Time zones complicate matters, while historical time zone mergers and daylight saving
time cause the timeline to be non-continuous, containing both gaps, as well as timestamps that occur twice.

This paper proposes a calendar designed from the ground up using elements that follow directly from nature. Each planet or colony has its own instance, with years and days having a
variable duration in order to accurately match true solar years and days. This allows people to match their daily cycle with their circadian rhythm, regardless of location.
Time of day is expressed as a proportion, rather than a unit, allowing people to keep track of exact proportions of time. Additionally, weeks and time of day are structured to match
the decimal system of human society, with this paper providing various linguistic hints for using the 10-day week, as well as the 10,000-beep daily clock.

The computational implementation discussed in this paper is publicly available and maps the calendar mechanism onto the discrete millisecond timeline used by computers. It resolves the duality between
a proportion of a day and a discrete millisecond and implements years, days and individual milliseconds in a mutually consistent manner. It includes an instance for planet Earth,
as well as a provisional instance for planet Mars, which can be adjusted once the site of the initial settlement is established.

The universal nature of the Lukashian calendar offers future human settlements on Mars and elsewhere a fresh start, away from the complexities of traditional calendar mechanisms,
toward a calendar that is simple and logical. Furthermore, this paper provides a starting point for further analog research using the Earth instance, listing various possible experiments and comparative
studies.

\section*{CRediT authorship contribution statement}
\textbf{Gert-Jan Schouten:} Conceptualization, Methodology, Software, Validation, Formal analysis, Investigation, Resources, Data Curation, Writing - Original Draft,
Writing - Review \& Editing, Visualization, Project administration.

\textbf{Michael Beye:} Writing - Review \& Editing.

\section*{Declaration of competing interest}
The authors declare that they have no known competing financial interests or personal relationships that could have appeared to influence the work reported in this paper.

\section*{Funding}
This research did not receive any specific grant from funding agencies in the public, commercial, or not-for-profit sectors.

\section*{Acknowledgments}
Analogous to Gangale's Darian calendar, the name “Lukashian” derives from Luke and Azshara (“Ash”), the children of the main author.

\section*{Public availability}
The source code of the reference implementation is available on GitHub~\cite{GITHUB} and the Java library is available on Maven Central~\cite{MAVEN}. It is published
under an Open Source license (included in the source code) and provides a calendar API with extensive functionality for calendar calculations. Instances for planets
Earth and Mars are included.

\bibliographystyle{elsarticle-num}
\bibliography{paper}

\end{document}